\documentclass[a4paper,11pt]{article}
\usepackage{jinstpub} 
\usepackage{lineno}
\usepackage{nicefrac}
\usepackage{graphicx}
\usepackage{caption}
\usepackage{subcaption}

\title{\boldmath Design and optimisation of radiation resistant AC- and DC-coupled resistive LGADs}

\author[a,b,1]{A.~Fondacci,\note{Corresponding author.}}
\author[b]{T.~Croci,}
\author[c,b]{D.~Passeri,}
\author[d,e]{R.~Arcidiacono,}
\author[e]{N.~Cartiglia,}
\author[f]{M.~Boscardin,}
\author[f]{M.~Centis Vignali,}
\author[f]{G.~Paternoster,}
\author[f]{O.~Hammad Ali,}
\author[g,e]{L.~Lanteri,}
\author[d,e]{L.~Menzio,}
\author[e]{F.~Siviero,}
\author[e]{M.~Ferrero,}
\author[g,e]{V.~Sola,}
\author[b]{A.~Morozzi,}
\author[h,b]{F.~Moscatelli,}
\affiliation[a]{Dipartimento di Fisica, Università degli Studi di Perugia,\\
Via Alessandro Pascoli, 06123, Perugia, Italy}
\affiliation[b]{Istituto Nazionale di Fisica Nucleare (INFN) - Sezione di Perugia,\\
Via Alessandro Pascoli, 06123, Perugia, Italy}
\affiliation[c]{Dipartimento di Ingegneria, Università degli Studi di Perugia,\\
Via Goffredo Duranti, 93, 06125, Perugia, Italy}
\affiliation[d]{Università del Piemonte Orientale,\\
Largo Donegani, 2, 28100, Novara, Italy}
\affiliation[e]{Istituto Nazionale di Fisica Nucleare (INFN) - Sezione di Torino,\\
Via Pietro Giuria, 1, 10125, Torino, Italy}
\affiliation[f]{Fondazione Bruno Kessler (FBK),\\
Via Sommarive, 18, 38123, Trento, Italy}
\affiliation[g]{Dipartimento di Fisica, Università degli Studi di Torino,\\
Via Pietro Giuria, 1, 10125, Torino, Italy}
\affiliation[a]{CNR - Istituto Officina dei Materiali (IOM),\\
Via Alessandro Pascoli, 06123, Perugia, Italy}

\emailAdd{alessandro.fondacci@pg.infn.it}

\abstract{Future high-energy physics experiments require a paradigm shift in radiation detector design. In response to this challenge, resistive LGADs that combine Low Gain Avalanche Diode technology with resistive readout have been developed. The prototypes created so far, employing AC-coupled contacts, have demonstrated impressive performance, achieving a temporal resolution of 38 ps and a spatial resolution of 15 µm with a pixel pitch of 450 µm.

To tackle some of the issues encountered up to this point, particularly the non-uniform response across the entire surface of the detector, a new version with DC-coupled contacts has recently been developed. The Synopsys\textsuperscript \textregistered\ Sentaurus TCAD simulations that have guided the design of their first production, released by the Fondazione Bruno Kessler in November 2024, will be presented below along with a concise summary of the history of the prototypes with AC-coupled contacts.}

\keywords{Solid state detectors, Particle tracking detectors, Timing detectors, Radiation damage to detector materials, Detector modelling and simulations II (electric fields, charge transport, multiplication and induction, pulse formation, electron emission, etc).}

\begin{document}
\maketitle
\flushbottom

\section{Introduction}
\label{sec:Intorduction}

    Silicon detectors have driven discoveries in high-energy physics (HEP) experiments for several decades~\cite{HARTMANN201225}. Indeed, leveraging standard photolithography techniques, they can be easily segmented to achieve the desired spatial resolution, which can be expressed as $\sigma_s \simeq \text{Pitch}/\sqrt{12}$\footnote{Formula valid for a digital position-finding algorithm in which the impinging position is determined by the coordinates of the pixel with the highest signal.}. Moreover, with the introduction of Low Gain Avalanche Diode (LGAD) technology~\cite{PELLEGRINI201412}, silicon detectors have also gained a reputation for excellent timing resolution, leading to the emergence of the blossoming field of 4D-tracking, which refers to the concurrent tracking of charged particles in space and time.

    However, pushed segmentation and LGAD technology alone may not fulfil all the requirements of future HEP experiments~\cite{Detector:2784893}, which demand spatial resolutions of just a few microns, timing resolutions of tens of picoseconds, low power consumption, and a reduced material budget. Decreasing the pixel size would increase the number of tracker channels, raising power consumption. A thick cooling infrastructure would also be necessary, potentially violating the low material budget specification. 

    Therefore, a new radiation detector architecture that combines LGAD technology with resistive readout, i.e., the resistive LGAD~\cite{Cartiglia2015}, has been introduced to meet the challenging requirements of future HEP experiments. The following sections will briefly review the performance achieved with AC-coupled electrode prototypes (RSDs) and summarise the TCAD (Technology Computer-Aided Design) simulation outcomes that guided the first production of the latest evolution with DC-coupled electrodes (DC-RSDs), released by the Fondazione Bruno Kessler (FBK) in November 2024. The analysis will primarily focus on spatial and timing resolutions, while other figures of merit, such as detection rate, will not be discussed in this work.

\section{AC-coupled resistive LGADs}
\label{sec:AC-RSD}

\begin{figure}[h]
        \centering
        \begin{subfigure}{0.49\textwidth}
            \centering
            \includegraphics[width=\textwidth]{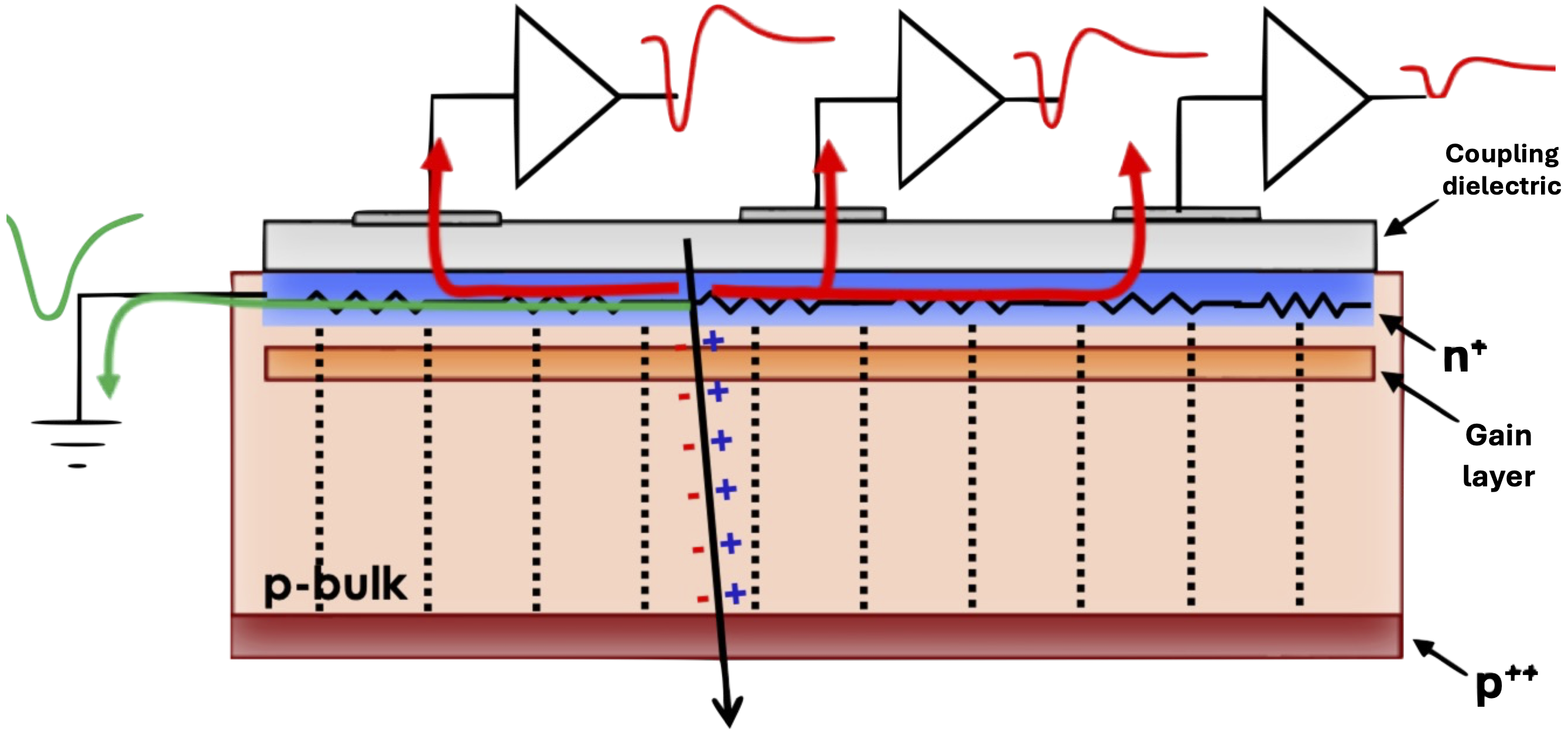}
            \caption{RSD}
            \label{Figure1}
        \end{subfigure}
        \hfill
        \begin{subfigure}{0.49\textwidth}
            \centering
            \includegraphics[width=\textwidth]{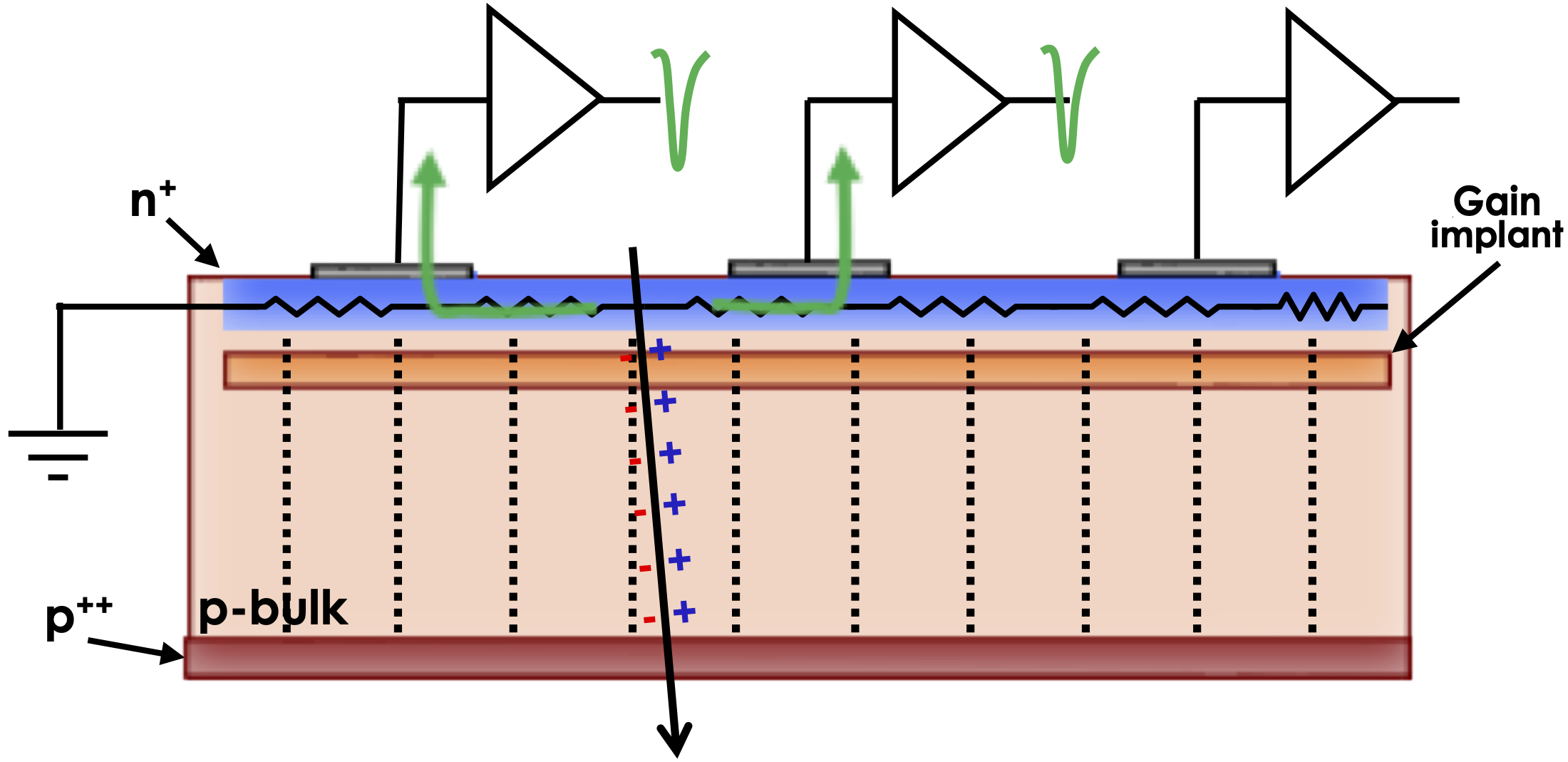}
            \caption{DC-RSD}
            \label{Figure2}
        \end{subfigure}
        \caption{Cross-sections of an AC-coupled resistive LGAD~(a) and a DC-coupled resistive LGAD~(b), illustrating how the signal flows after the transit of a particle.}
    \end{figure}

    In an RSD, individual n-wells are replaced by a single resistive layer, beneath which the gain implant is uniform (Fig.~\ref{Figure1}), thereby ensuring a 100\% fill factor. When a particle hits the detector, primary electron-hole pairs are generated, along with secondary pairs from avalanche multiplication within the gain layer. These charges induce a signal in the resistive plane that propagates from the point of impact toward the ground. The latter is represented by the AC-coupled electrodes for the fast components of the signal, and therefore, the signal splits among the nearby  AC-coupled pads like a current in an impedance divider, where the impedance is that of the paths connecting the impact point to each of the electrodes. Signal sharing is thus an inherent characteristic of the detector, which allows for the analog reconstruction of the impact position when coupled with the internal signal amplification provided by LGAD technology. RSDs can thus achieve much better spatial resolutions compared to those possible with binary readout or the same ones but with larger pixels, therefore allowing for a reduction in the number of tracker channels. Furthermore, due to internal signal amplification, the substrate thickness can be minimised to adhere to low material budget specifications, while still producing the large signal amplitudes necessary for timing.

    In 2019 and 2021, FBK released two different productions of RSDs. The first one, RSD1~\cite{TORNAGO2021165319}, aimed to demonstrate the proof of concept and involved structures with square pads of varying sizes arranged in a square matrix with different pitches. Intensive characterisation using lasers and charged particles showed that these prototypes could achieve timing resolutions typical of LGADs, around 40 ps, and spatial resolutions of up to 5 µm with an interpad distance of 100 µm. This confirmed that excellent spatial resolutions could be attained even with large pixels, allowing for a reduction in the number of tracker channels.

    The impressive performances observed in the RSD1 production prototypes were explicitly related to the characterisation of an optimal sub-area of the pixel—the central part where the signal is shared among all four pads rather than just between two or exclusively being collected by a single pad. To achieve such excellent performance over the entire pixel area, the RSD2~\cite{ARCIDIACONO2023168671} production was developed, focusing on alternative electrode geometries. Notably, the cross-shaped pads performed exceptionally well across the entire pixel surface since they minimise the area where signal sharing does not involve all four electrodes of the hit pixel. The RSD2 production thus confirmed the feasibility of achieving a timing resolution of 38 ps and a spatial resolution below 5\% of the pitch, such as 15 µm with a $450 \times 450$ µm$^2$ pixel, which scales linearly with it.

\section{DC-coupled resistive LGADs}
\label{sec:DC-RSD}

    In resistive LGADs, the pixel size can thus be determined by the expected detector occupancy rather than by the desired spatial resolution. This makes it crucial to confine signal sharing to the pads of the hit pixel to prevent performance loss due to crosstalk between pixels. Additionally, this will ensure uniformity in response across the entire sensor surface, allowing all pixels to behave the same. However, in RSDs, only the high-frequency components of the signal are collected by the AC-coupled electrodes, leading to an average signal leak of approximately 30\% from the hit pixel, as has been experimentally observed~\cite{MENZIO2024169526}. A new version of resistive LGADs with DC-coupled contacts (Fig.~\ref{Figure2})~\cite{MENZIO2022167374} has thus been developed to address this issue. Full 3D TCAD simulations strongly supported the design phase of this latest release, and a detailed description of the simulation setup can be found in~\cite{10411098}.

    Building on the success of crosses with RSDs~\cite{ARCIDIACONO2023168671}, initial attempts have been made to confine the signal using DC-coupled cross-shaped pads by evaluating different lengths for their arms~\cite{MOSCATELLI2024169380}. It was observed that when the arms are a significant fraction of the pitch, nearly the entire signal (97\%) is contained within the hit pixel. Therefore, closely surrounding the pixel with electrodes is an effective strategy to limit signal sharing, at least when the contact resistance is low. However, when contact resistance increases, the signal continues flowing in the resistive plane rather than being captured by the readout electrodes.

    \begin{figure}[h]
        \centering
        \begin{subfigure}{0.45\textwidth}
            \centering
            \includegraphics[width=\textwidth]{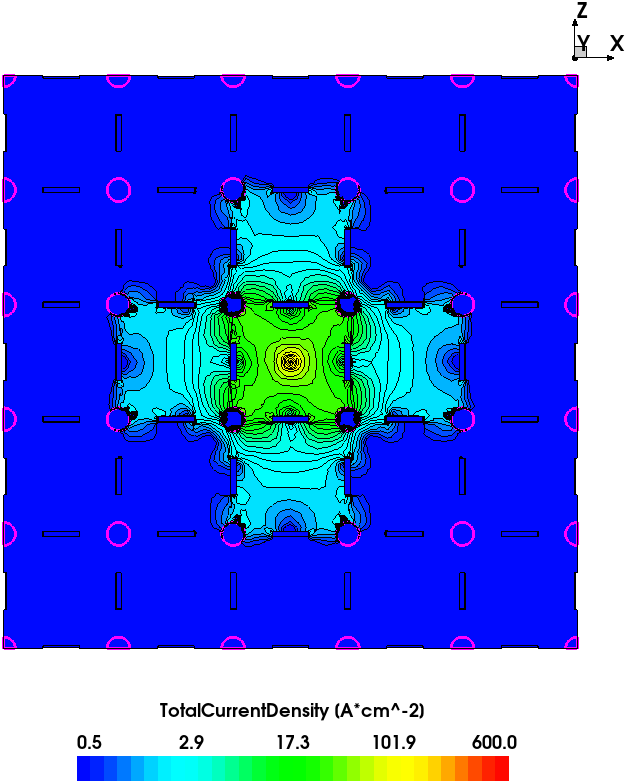}
            \caption{Trenches covering 40\% of pad gap}
        \end{subfigure}
        \hfill
        \begin{subfigure}{0.45\textwidth}
            \centering
            \includegraphics[width=\textwidth]{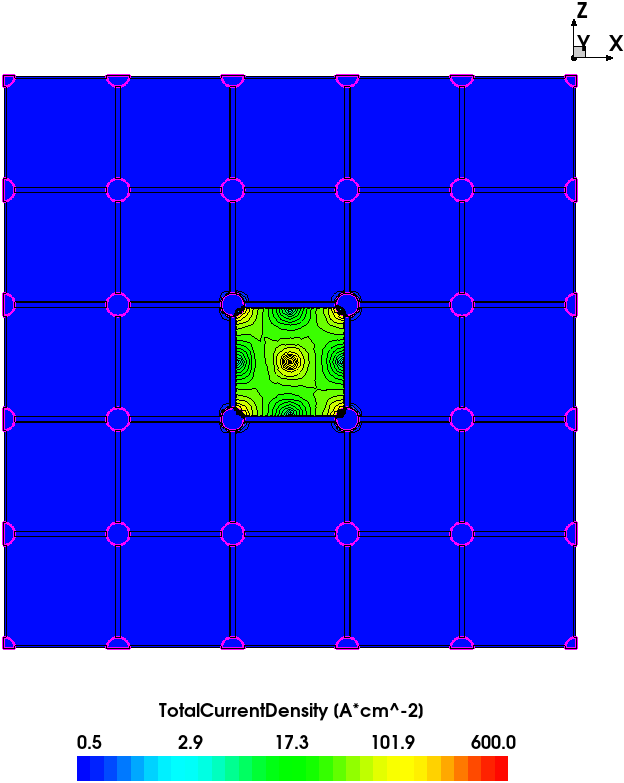}
            \caption{Pad to pad trenches}
        \end{subfigure}
        \caption{Signal spreading in the resistive plane as a current density map for two silicon oxide trench extensions a few tens of picoseconds after a minimum ionising particle (MIP) was shot in the centre pixel. To avoid crosstalk between pixels, i.e., to confine the signal entirely within the hit pixel, the trenches have to extend from pad to pad.}
        \label{Figure3}
    \end{figure}

    While cross electrodes with long arms and low contact resistance effectively confine the signal, they, unfortunately, introduce distortion errors in reconstructing the impact position. In these detectors, the latter can be estimated using a charge centre algorithm, which weighs the charge collected by each electrode based on its centre position~\cite{10411098}. For example, if a charged particle hits one of the tips of a cross, the signal will be collected by that electrode alone, leading to the reconstruction of the impact position at the centre of the electrode, even though the hit occurred at the tip. The reconstruction error will, therefore, correspond to the length of one arm of the cross, typically on the order of hundreds of microns. It is therefore necessary to use small, ideally point-like electrodes to mitigate this effect, although such electrodes alone cannot confine the signal.

    By leveraging the silicon oxide trench technology already implemented in Silicon Photomultipliers (SiPM) or trench-isolated LGADs~\cite{PATERNOSTER2021164840}, it is possible to achieve good signal confinement even if the pads are circular and small. By introducing a trench between each pair of electrodes interrupts the resistive plane, thereby isolating the single pixels and ensuring that all signal is collected by the four pads of the hit pixel. TCAD simulations have demonstrated that it is crucial for the trench to extend from pad to pad; otherwise, there is a risk of the signal escaping from the hit pixel (Fig.~\ref{Figure3}).

    The issue of contact resistance has also been investigated for structures with small circular pads and pad-to-pad silicon oxide trenches. When the contact resistance is high, the signal that reaches the four electrodes at the corners of the hit pixel tends to bypass the readout pads and flow into adjacent pixels. This confirms the critical importance of achieving low contact resistance in resistive LGADs.

\begin{figure}[h]
        \centering
        \includegraphics[width=\textwidth]{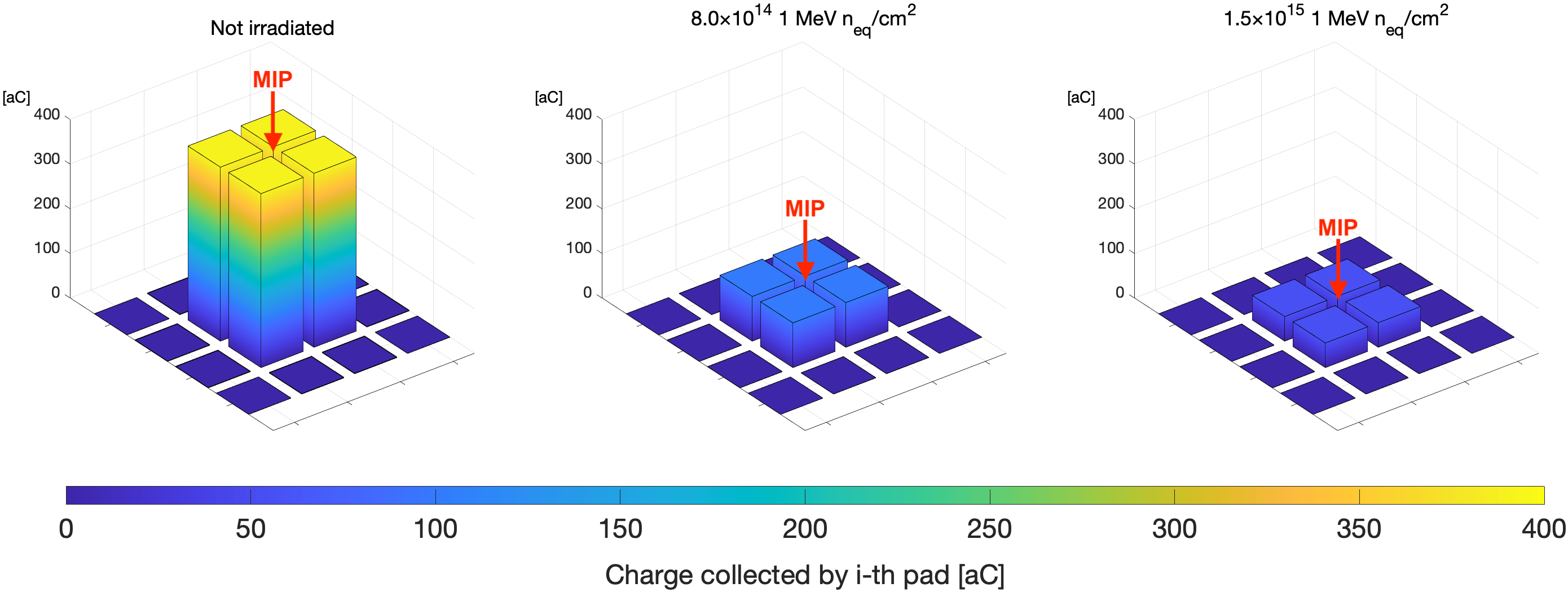}
        \caption{Charge collected from the electrodes of a DC-RSD with pad-to-pad silicon oxide trenches for three different fluences. A MIP was shot in the centre pixel.}
        \label{Figure4}
    \end{figure}

    Lastly, still for the structure incorporating pad-to-pad trenches, the behaviour after irradiation has also been examined. The latest version of the Perugia radiation damage model~\cite{Morozzi:2024KG} was employed for this investigation. Figure~\ref{Figure4} shows the charge collected by the pads for three different fluences when a MIP hits the centre pixel. Although the overall collected charge decreases with increasing fluence due to charge trapping and acceptor removal, the percentage of charge collected by each pad remains nearly constant regardless of the irradiation level. This indicates that the behaviour of the resistive plane remains almost unchanged, as it is an implant with a high peak concentration and a small depth, making it only slightly affected by donor removal as fluence increases.
    
\section{Conclusion}
\label{sec:Conclusion}

    Resistive LGADs are a promising solution to meet the requirements of future high-energy physics experiments. Prototypes developed so far, with AC-coupled contacts (RSDs), have demonstrated exceptional spatial and temporal resolutions even with large pixel sizes. For instance, structures with a 450 µm pitch and cross-shaped electrodes have achieved a temporal resolution of 38 ps and a spatial resolution of 15 µm. As a result, the number of tracker channels can be reduced, leading to lower power consumption.

    An improved version with DC-coupled contacts (DC-RSDs) has been designed to standardise the response across the entire detector surface, aiming to limit signal sharing exclusively to the pads of the hit pixel. Full 3D TCAD simulations strongly supported the design of their first production, released in November 2024 by FBK. These simulations revealed that surrounding the pixels with pads is effective for confining signal spread, but it also introduces distortion in reconstructing the impact position. As a solution, small circular pads have been chosen to minimise reconstruction errors, complemented by pad-to-pad silicon oxide trenches to ensure signal containment within the hit pixel. Moreover, it is crucial to maintain low contact resistance so as not to lose signal confinement. Lastly, it has also been observed that irradiation does not significantly affect the properties of the resistive plane, which is fundamental to the operation of resistive LGADs, as it is slightly affected by donor removal.

    The characterisation of these new devices has already started. It includes laboratory tests and a test beam at DESY in December 2024, with another test beam scheduled for March 2025. The results from these tests will assess the devices' performance and validate the robustness of the simulation framework used in their design.

\acknowledgments

    This work has received funding from INFN CSN5 through the 4DSHARE research project, PRIN MIUR projects 2017L2XKTJ ‘4DInSiDe’ and 2022KLK4LB ‘4DSHARE’, Compagnia San Paolo (TRAPEZIO grant), European Union's Horizon Europe research and innovation program under grant agreement no. 101057511. We acknowledge the RD50 and DRD3 collaborations, CERN.

\bibliographystyle{JHEP}
\bibliography{biblio.bib}

\end{document}